\documentclass[twocolumn,preprintnumbers,amsmath,superscriptaddress,amssymb,aps]{revtex4-1}

\usepackage{graphicx}% Include figure files
\usepackage{dcolumn}% Align table columns on decimal point
\usepackage{bm}% bold math
\usepackage{amsmath}
\usepackage{amssymb}
\usepackage{graphicx}
\usepackage{indentfirst}
\usepackage{booktabs}
\usepackage{multirow}
\usepackage{colortbl}

\selectfont

\begin{document}
\title{Coexisting Triferroic and Multiple Types of Valley Polarization by Structural Phase Transition in Two-Dimensional Materials}

\author{Chao Wu}
\address{State Key Laboratory for Mechanical Behavior of Materials, School of Materials Science and Engineering, Xi'an Jiaotong University, Xi'an, Shaanxi, 710049, People's Republic of China}
\author{Hanbo Sun}
\address{State Key Laboratory for Mechanical Behavior of Materials, School of Materials Science and Engineering, Xi'an Jiaotong University, Xi'an, Shaanxi, 710049, People's Republic of China}
\author{Pengqiang Dong}
\address{State Key Laboratory for Mechanical Behavior of Materials, School of Materials Science and Engineering, Xi'an Jiaotong University, Xi'an, Shaanxi, 710049, People's Republic of China}
\author{Yin-Zhong Wu}
\address{School of Physical Science and Technology, Suzhou University of Science and Technology, Suzhou 215009, China }
\author{Ping Li}
\email{pli@xjtu.edu.cn}
\address{State Key Laboratory for Mechanical Behavior of Materials, School of Materials Science and Engineering, Xi'an Jiaotong University, Xi'an, Shaanxi, 710049, People's Republic of China}
\address{State Key Laboratory of Silicon and Advanced Semiconductor Materials, Zhejiang University, Hangzhou, 310027, People's Republic of China}
\address{State Key Laboratory for Surface Physics and Department of Physics, Fudan University, Shanghai, 200433, People's Republic of China}

\date{\today}

\begin{abstract}
The multiferroic materials, which coexist magnetism, ferroelectric, and ferrovalley, have broad practical application prospects in promoting the miniaturization and integration of spintronic and valleytronic devices. However, it is rare that there are triferroic orders and multiple types of valley polarization in a real material. Here, we propose a mechanism to realize triferroic order coexistence and multiple types of valley polarization by structural phase transition in two-dimensional (2D) materials. The 1T and 2H phase OsBr$_2$ monolayers exhibit non-magnetic semiconductor and ferromagnetic semiconductor with valley polarization up to 175.49 meV, respectively. Interestingly, the 1T phase OsBr$_2$ bilayer shows the tri-state valley polarization due to lattice symmetry breaking, while the valley polarization of 2H phase bilayer originates from the combined effect of time-reversal symmetry breaking and spin-orbit coupling. Furthermore, the valley polarization and ferroelectric polarization of 1T phase AB stackings and 2H phase AA stackings can be manipulated via interlayer sliding. Importantly, we have verified that the 2H phase can be transformed to 1T phase by Li$^+$ ion intercalation, while the 2H phase can occur the structural phase transition into the 1T phase by infrared laser induction. Our work provides a feasible strategy for manipulating valley polarization and a design idea for nano-devices with nonvolatile multiferroic properties.
\end{abstract}

\maketitle
\section{Introduction}
The valley, as the local energy extremum point of a band structure in momentum space \cite{1,2,3,4}, provides a valley degree of freedom for information storage analogous to charge and spin in spintronics. The valleys are generally degenerate, since certain $\emph{k}$ points can be connected by specific symmetry, such as inversion symmetry ($\hat{P}$) and time-reversal symmetry ($\hat{T}$). In order to use the valley indicator as an information carrier, it is requisite to manipulate the carriers in the valley to generate valley polarization. There are many approaches that can be applied to eliminate valley degeneracy, such as magnetic field \cite{5,6}, magnetic doping \cite{7}, magnetic proximity effect \cite{8,9}, electric field \cite{10}, and optical pumping \cite{11,12}. However, the valley polarization induced by these methods is transient or volatile, which is difficult to implement for practical applications.

In order to facilitate the application of valleytronics, it is demanded that valley polarization is non-volatile and can be switched between equivalent valleys, i.e. ferrovalley \cite{13}. Two main approaches to proposed be to realize ferrovalley. One approach is to break the $\hat{P}$ symmetry with ferroelectricity, which the valley polarization appears simultaneously with ferroelectric polarization \cite{14,15,16}. More importantly, the valley polarization direction and ferroelectric polarization direction are locked together, where are easily switched by an external electric field. The other approach is to break the $\hat{T}$ symmetry with magnetism, which can be employed in two-dimensional (2D) valley polarized semiconductor with the intrinsic magnetism \cite{17,18,19,20,21}. Such as VSe$_2$ \cite{13}, CrOBr \cite{22}, which only realize single valley polarization. Moreover, the multiferroic properties of these materials are ferromagnetic and ferrovalley, which are lack of ferroelectric. In order for these materials to be ferroelectric property, they must form heterostructures with ferroelectric materials \cite{23}. However, it is challenging to realize two approaches of valley polarization in the same material system. It indicates that it is urgent to investigate new mechanisms for ferrovalley. And realizing triferroic order including the magnetic, ferroelectric, and ferrovalley coexist. The coexistence of triferroic orders can enrich the controllable physical properties and broaden the application scenarios of materials.

Recently, the unconventional superconductivity \cite{24}, Mott insulation state \cite{25}, magnetism \cite{26}, quantum anomalous Hall effect \cite{27}, and other novel physical properties \cite{28,29} were discovered in the magic-angle graphene system, indicating that stacking of 2D materials could effectively regulate the physical properties. For example, interlayer sliding can induce ferroelectricity \cite{30,31,32,33}, bulk photovoltaic effect \cite{34}, magnetic phase transition \cite{20,35}, and altermagnetism \cite{36}. Moreover, Li$^+$ ion intercalation and infrared laser can induce the intertransition between 1T and 2H phases \cite{37,38}. Whether it is possible to realize multiple valley polarization and multiple ferroic orders coupling through structural phase transitions?

In this work, we propose a novel mechanism for realizing triferroic and multiple types of valley polarization. This mechanism can obtain the coupling between magnetism, ferroelectric, and ferrovalley by the sliding of the bilayer. Moreover, the multiple types of valley polarization can also be achieved by structural phase transition induced symmetry breaking. Furthermore, based on the density functional theory (DFT), we also exhibit that this mechanism can be verified in a real material of bilayer OsBr$_2$. We found that the magnetism, ferroelectric, and valley polarization can be switched by tuning the sliding of bilayer OsBr$_2$. The structural phase transition from the 2H phase to the 1T phase of OsBr$_2$ was induced by Li$^+$ ion intercalation, realizing the multiple types of valley polarization. Importantly, we reveal the microscopic mechanism of valley polarization by Hamiltonian and symmetry. Our study not only creates a pathway for triferroic and multiple types of valley polarization, but also extends the valley polarization.

\section{STRUCTURES AND COMPUTATIONAL METHODS}
All first-principles calculations based on the framework of the DFT are implemented the Vienna $Ab$ $initio$ Simulation Package (VASP) \cite{39,40,41}. The exchange-correlation energy is described by the generalized gradient approximation (GGA) with the Perdew-Burke-Ernzerhof (PBE) \cite{42}. The plane-wave basis with a kinetic energy cutoff is set to be 500 eV. The $k$ meshes of Brillouin zone is adopted using $18\times 18\times 1$ $\Gamma$-centered. A vacuum of 20 $\rm \AA$ is added along the c-axis, to avoid the interaction between the sheet and its periodic images. The total energy and force convergence criterion are set to be 10$^{-6}$ eV and -0.001 eV/$\rm \AA$, respectively. The zero damping DFT-D3 method of Grimme is considered for van the vdW correction in bilayer OsBr$_2$ \cite{43}. To describe strongly correlated 5d electrons of Os \cite{44}, the GGA + U method is applied. We find that the result of U = 2 eV is the closest to the result of HSE06, as shown in Figure S1 and Figure S2. Therefore, the Coulomb repulsion U is set to 2.0 eV. To investigate the dynamical stability, the phonon spectra are calculated by the PHONOPY code using a $4\times 4\times 1$ supercell \cite{45}. In addition, the sliding barriers and ferroelectric switching path are calculated using climbing image nudged elastic band method \cite{46}. The maximally localized Wannier functions (MLWFs) are used to construct an effective tight-binding Hamiltonian to study the Berry curvature \cite{47}. The Berry curvature calculations used the formula
\begin{equation}
	\Omega^z(\textbf{k})=-\sum_{n}f_{n}\sum_{n\prime \neq n}\frac{2Im \left \langle \psi_{nk} \mid v_{x} \mid \psi_{n\prime k} \right \rangle \left \langle \psi_{n\prime k} \mid v_{y} \mid \psi_{nk} \right \rangle}{(E_{n\prime}-E_{n})^2},
\end{equation}
where $\Omega^z(\textbf{k})$ is the Berry curvature in the reciprocal space, $v_{x}$ and $v_{y}$ are operator components along the x and y directions, and $f_{n}=1$ for the occupied bands, respectively \cite{48,49,50}.

\section{RESULTS AND DISCUSSION }	
\subsection{The Model of Coexisting Triferroic and Multiple Types of Valley Polarization}
%%%%%%%%%%%%%%%%%%%%%%%%%%%%%%%%%%%%%%%%%%%%%%%%%%%%%%%%%%%%%%%%%%
\begin{figure}[htb]
\begin{center}
\includegraphics[angle=0,width=1.0\linewidth]{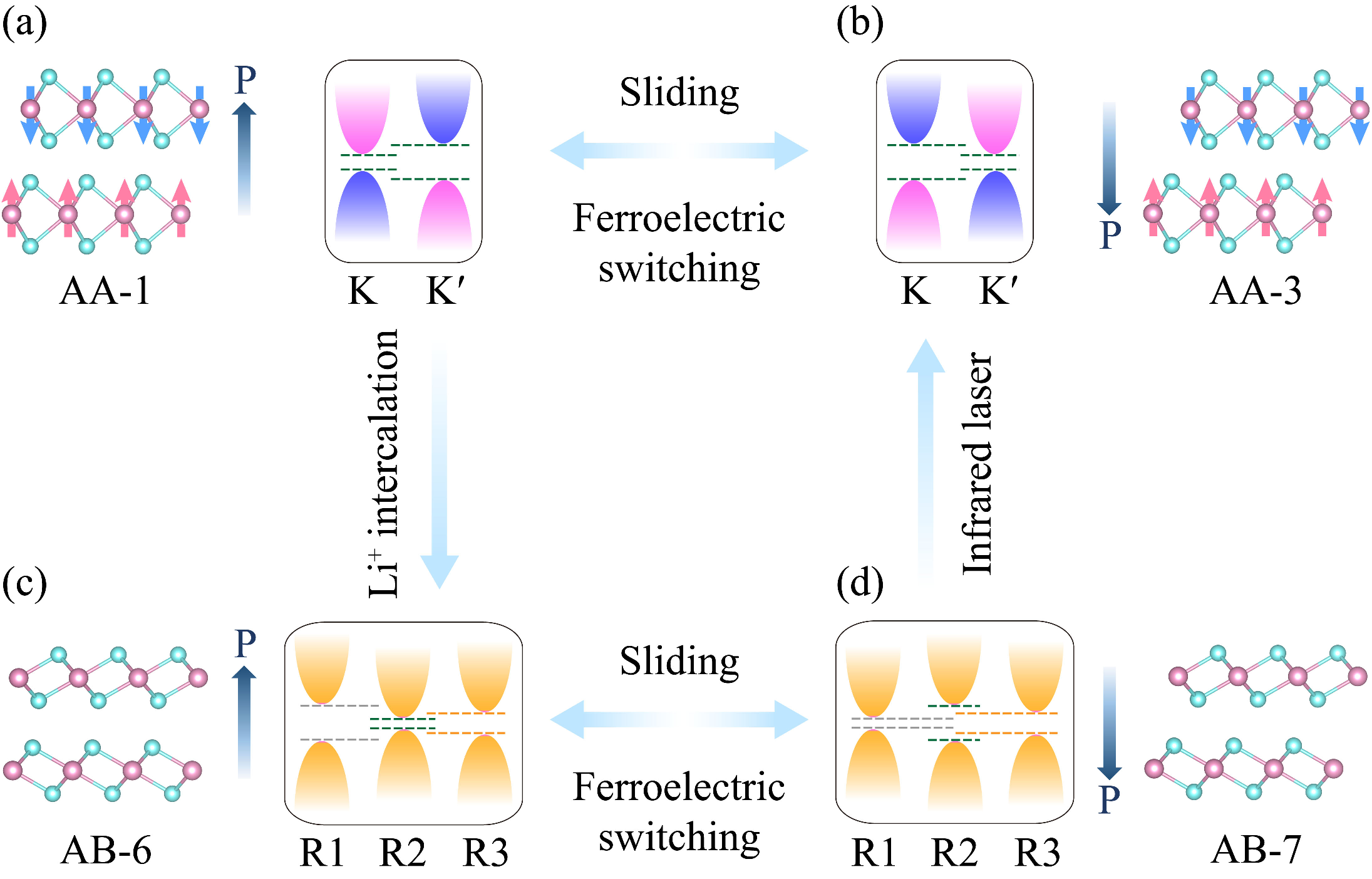}
\caption{Schematic of the coexisting triferroic: ferroelectric, magnetism, and ferrovalley, and multiple types of valley polarization. (a) The magnetic ground state of 2H phase AA-1 stacking bilayer lattice is AFM state, and exhibits spontaneous ferroelectric polarization and valley polarization. The AA-1 stacking slide to AA-3 stacking (b), the valley polarization can be manipulated via magnetization reversing and ferroelectric switching in AA-1 stacking (a) and AA-3 stacking (b). The 2H phase bilayer can be induced into 1T phase bilayer by Li intercalation. (c) 1T phase AB-6 stacking is non-magnetic, and also has spontaneous ferroelectric polarization and polymorphic valley polarization. The AB-6 stacking slide to AB-7, the valley polarization can also be controlled via ferroelectric switching in AB-6 stacking (c) and AB-7 stacking (d). The 1T phase bilayer can be translated to the 2H phase bilayer by the infrared laser. Magenta, blue and orange cones denote  spin up, spin down, and nonmagnetic bands, respectively. Dark gray arrows represent ferroelectric polarization $\emph{P}$.
}
\end{center}
\end{figure}
%%%%%%%%%%%%%%%%%%%%%%%%%%%%%%%%%%%%%%%%%%%%%%%%%%%%%%%%%%%%%%%%%%

The 2D triferroic systems with the coexistence of magnetism, ferrovalley, and ferroelectric (see Figure S3) and multiple types of valley polarization are established. Actually, for a monolayer system with intrinsically spontaneous valley polarization originating from the $\hat{T}$ symmetry breaking, it can realize triferroic by bilayer stacking and interlay sliding. Figure 1 shows realizing the mechanism of 2D triferroic and multiple types of valley polarization. In 2H phase AA-1 stacking, it exhibits antiferromagnetic (AFM) ground state and +z axis direction out-of-plane ferroelectric polarization (signed by dark gray arrows). Simultaneously, it has spontaneous valley polarization due to the $\hat{P}$ and $\hat{T}$ symmetry breaking. When the AA-1 stacking slide to AA-3 stacking, leading to switch the direction of ferroelectric polarization and valley polarization. Moreover, the phase transition from 2H to 1T phase can be induced by Li$^+$ ion intercalation. It's worth noting that the 1T phase magnetism disappears. Interestingly, it exhibits tri-state valley polarization, and still shows the +z axis direction out-of-plane ferroelectric polarization. As the AB-6 stacking slides to AB-7 stacking, similarly, the ferroelectric polarization and valley polarization are switched.

\subsection{Sliding Energy Barriers, Ferroelectricity, Magnetism of Bilayer OsBr$_2$ }
Figure S4(a, e) shows the crystal structure of 1T and 2H phase monolayer OsBr$_2$. They have a hexagonal lattice and space group $\emph{P$\bar{3}$m1}$ for 1T phase, $\emph{P$\bar{6}$m2}$ for 2H phase, respectively. Each unit cell contains one Os atom and two Br atoms. The optimized lattice constant of 1T and 2H phases are 3.81 $\rm \AA$ and 3.61 $\rm \AA$. For the 1T phase, each Os atom is surrounded by six Br atoms, forming a triangular prism. Therefore, the $\hat{P}$ symmetry is broken for 1T phase OsBr$_2$ monolayer. Moreover, 1T and 2H phases hold octahedral and triangular prism crystal fields, respectively. As shown in Figure S4(b, f), the d orbitals of 1T phase split into three t$_{\rm 2g}$ (d$_{\rm xy}$, d$_{\rm xz}$, d$_{\rm yz}$) and two e$_{\rm g}$ (d$_{\rm x^2-y^2}$, d$_{\rm z^2}$,) orbitals, while the d orbitals of 2H phase split into one a$_{\rm 1g}$ (d$_{\rm z^2}$), two e$_{\rm 1}$ (d$_{\rm xy}$, d$_{\rm x^2-y^2}$), and two e$_{\rm 2}$ (d$_{\rm xz}$, d$_{\rm yz}$). In the 1T phase, the 5d$^6$ electrons of Os hold the filled t$_{\rm 2g}$ majority and minority states. As a result, the system has no net magnetic moment and doesn't exhibit magnetism. On the contrary, the electrons hold the filled a$_{\rm 1g}$, e$_{\rm 1}$, and e$_{\rm 2}$ majority states and the half-filled e$_{\rm 1}$ minority state. The feature makes 2H phase OsBr$_2$ monolayer semiconducting material with the magnetic moment of 4$\mu_{\rm B}$ per Os.

To confirm the magnetic ground state of the 2H phase OsBr$_2$ monolayer, three possible magnetic configurations are considered in the $2\times 2\times 1$ supercell, namely non-magnetic (NM), ferromagnetic (FM), and AFM structures. The energy of the FM state is 4086.85 meV, and 865.96 meV lower than the NM and AFM states, respectively. It indicates that the FM is the magnetic ground state. Moreover, the magnetic anisotropy energy (MAE) is very important to confirm the thermal stability of magnetic ordering \cite{51}. Here, the MAE is defined as MAE = E$_{\rm 100}$ - E$_{\rm 001}$, where E$_{\rm 100}$ and E$_{\rm 001}$ denote the total energy of the magnetic moment along [100] and [001], respectively. The MAE is 41.17 meV, indicating the direction of easy magnetization along the z axis.

Next, we investigate the band structure and valley properties of 1T and 2H phase OsBr$_2$ monolayer. As shown in Figure S4(c), 1T phase OsBr$_2$ monolayer shows semiconductor property with a band gap of 1.90 eV and R1, R2, R3 valence and conduction valley degeneracy under the absence of spin-orbit coupling (SOC). Figure S4(g) exhibits the orbital-resolved band structure of 2H phase OsBr$_2$ monolayer without considering the SOC. The valence band maximum (VBM) is dominated by Os d$_{\rm xy}$/d$_{\rm x^2-y^2}$ orbitals, while the conduction band minimum (CBM) is mainly contributed by Os d$_{\rm z^2}$. Importantly, the K and K' valleys of VBM and CBM are energetic degeneracy. When the SOC is included, the valley degeneracy of the K and K' points disappears, as shown in Figure S4(h). The total valley splitting is 175.49 meV, where the valley splitting of the VBM and CBM are 142.19 meV and 33.30 meV, respectively. It is larger than VSe$_2$ (90 meV) \cite{13}, CrOBr ($\thicksim$ 112 meV) \cite{22}, YI$_2$ ($\thicksim$ 105 meV) \cite{52}, VSiGeP$_4$ ($\thicksim$ 100 meV) \cite{53}, Mn$_2$P$_2$S$_3$Se$_3$ \cite{54}, MoTe$_2$/EuO ($\thicksim$ 20 meV) \cite{55}, MoTe$_2$/CoCl$_2$ ($\thicksim$ 20 meV) \cite{56}, and other ferrovalley materials.

To understand the origin of valley splitting in the 2H phase OsBr$_2$ monolayer, we construct an effective Hamiltonian model using the SOC effect as a perturbation term,
\begin{equation}
\hat{H}_{SOC} = \lambda \hat{S} \cdot \hat{L} = \hat{H}_{SOC}^{0} + \hat{H}_{SOC}^{1},
\end{equation}
where $\hat{S}$ and $\hat{L}$ are spin angular and orbital angular operators, respectively.
$\hat{H}_{SOC}^{0}$ represents  the interaction between the same spin states, while $\hat{H}_{SOC}^{1}$ denotes the interaction between the opposite spin states. Since the VBM and CBM are both spin down bands, hence, the term $\hat{H}_{SOC}^{1}$ can be neglected. Moreover, the $\hat{H}_{SOC}^{0}$ term can be rewritten by the polar angles
\begin{equation}
\hat{H}_{SOC}^{0} = \lambda \hat{S}_{z'}(\hat{L}_zcos\theta + \frac{1}{2}\hat{L}_+e^{-i\phi}sin\theta + \frac{1}{2}\hat{L}_-e^{+i\phi}sin\theta),
\end{equation}
When the magnetization direction of 2H phase OsBr$_2$ monoalyer along the +z axis, $\theta$ = $\phi$ = 0$^ \circ$, the $\hat{H}_{SOC}^{0}$ term can be reduced as
\begin{equation}
\hat{H}_{SOC}^{0} = \lambda \hat{S}_{z} \hat{L}_z,
\end{equation}

It considers that the 2H phase has the C$_3$ symmetry and the orbital contributions of the VBM and CBM, we take $|$$\psi$$_v$$^{\tau}$$\rangle$=$\frac{1}{\sqrt{2}}$($|$d$_{xy}$$\rangle$+i$\tau$$|$d$_{x2-y2}$$\rangle$)$\otimes$$|$$\downarrow$$\rangle$, $|$$\psi$$_c$$^{\tau}$$\rangle$=$|$d$_{z^2}$$\rangle$$\otimes$$|$$\downarrow$$\rangle$ as the orbital basis for the VBM and CBM, where $\tau$ = $\pm$1 denote the valley index corresponding to $\rm K/\rm K'$. The energy of the K and K' valleys for the VBM and CBM can be written as E$_v$$^ \tau$ = $\langle$ $\psi$$_v$$^ \tau$ $|$ $\hat{H}$$_{SOC}^{0}$ $|$ $\psi$$_v$$^ \tau$ $\rangle$ and E$_c$$^ \tau$ = $\langle$ $\psi$$_c$$^ \tau$ $|$ $\hat{H}$$_{SOC}^{0}$ $|$ $\psi$$_c$$^ \tau$ $\rangle$, respectively. Then, the valley polarization of valence and conduction bands can be written as
\begin{equation}
E_{v}^{K} - E_{v}^{K'} = i \langle d_{xy} | \hat{H}_{SOC}^{0} | d_{x2-y2} \rangle - i \langle d_{x2-y2} | \hat{H}_{SOC}^{0} | d_{xy} \rangle \approx 4\alpha,
\end{equation}
\begin{equation}
E_{c}^{K} - E_{c}^{K'} = 0,
\end{equation}
where the $\hat{L}_z|d_{xy} \rangle$ = -2i$\hbar$$|d_{x2-y2} \rangle$, $\hat{L}_z|d_{x2-y2} \rangle$ = 2i$\hbar$$|d_{xy} \rangle$, and $\alpha = \lambda \langle d_{x2-y2} |\hat{S}_{z'}| d_{x2-y2} \rangle$. The analytical result confirms that the valley splitting for the valence band is consistent with our DFT calculations ($E_{v}^{K}$ - $E_{v}^{K'}$ = 142.19 meV). There is also some valley splitting in the CBM, which is due to the small number Os-d$_{\rm xz}$/d$_{\rm yz}$, and Br-p$_{\rm x}$/p$_{\rm y}$ orbits at the CBM.

The layer stacking of 2D materials may give rise to interesting physical phenomena that are different from those of monolayer materials \cite{20,52,53,57}. Here, we explore the unique physical behaviors of 1T and 2H phase OsBr$_2$ bilayer by stacking orders. As shown in Figure S5, we take AA-0 and AA-1 stackings, which are the lowest energy of 1T and 2H phase OsBr$_2$ AA stacking, as an example to explore electron localization function (ELF) and dynamic stability. The electrons are mainly concentrated around the atoms, while the electrons between the atoms are negligible, which indicates that all bonds are ionic bonding. Figure S5(c, f) shows the lack of imaginary frequency along the entire high-symmetry lines, meaning the dynamical stability for the 1T and 2H phase OsBr$_2$ bilayer. Moreover, the typical stacking configurations are shown in Figure S6 and Figure S7. The AA stacking OsBr$_2$ bilayer is built by placing one layer on top of the other, while AB stacking OsBr$_2$ bilayer can be obtained by rotating the upper layer 180$^\circ$ of AA stacking about the xy plane. Other stacking configurations can be acquired by AA or AB stacking sliding.

%%%%%%%%%%%%%%%%%%%%%%%%%%%%%%%%%%%%%%%%%%%%%%%%%%%%%%%%%%%%%%%%%%
\begin{figure}[htb]
\begin{center}
\includegraphics[angle=0,width=1.2\linewidth]{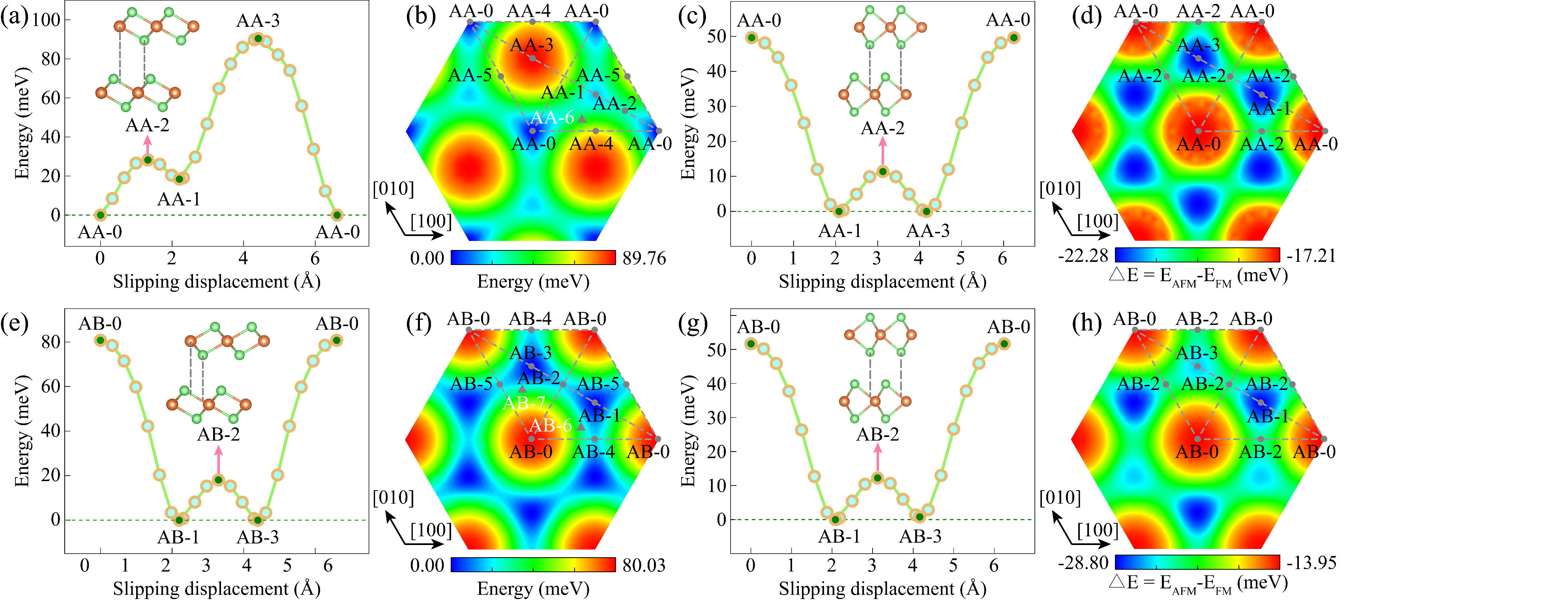}
\caption{Energy barriers via interlayer sliding in 1T and 2H phases OsBr$_2$ bilayers. (a, e) Sliding energy barrier along [$\bar{1}$10] direction for (a) 1T phase AA stacking and (e) AB stacking. The all space of lateral shifts for (b) 1T phase AA stacking and (f) AB stacking bilayer OsBr$_2$. (c, g) Sliding energy barrier along [$\bar{1}$10] direction for (c) 2H phase AA stacking and (g) AB stacking. The energy difference between interlayer AFM and FM states as a function of interlayer translation for the all space of lateral shifts (d) 2H phase AA stacking and (h) AB stacking OsBr$_2$.
}
\end{center}
\end{figure}
%%%%%%%%%%%%%%%%%%%%%%%%%%%%%%%%%%%%%%%%%%%%%%%%%%%%%%%%%%%%%%%%%%

The sliding energy barriers of 1T and 2H phase AA and AB stacking configurations are explored to acquire the most stable stacking pattern. In the 1T phase AA stackings, AA-3 stacking exhibits the highest energy, while AA-0 stacking shows the lowest energy, as shown in Figure 2(a). The total energy of AA-1 stacking is 18.42 meV higher than AA-0 stacking, indicating that AA-1 stacking is metastable. Moreover, the energy barrier of 28.22 meV requires overcoming from AA-0 stacking sliding to AA-1 stacking. For the 1T phase AB stackings [see Figure 2(e)], AB-1 and AB-3 stackings show the lowest energy, while the AB-0 stacking exhibits the highest energy. The energy barrier for the AB-1 stacking sliding to AB-3 stacking is 18.13 meV. More interestingly, the energy barrier of 2H phase AA stacking is similar to that of 1T phase AB stacking [see Figure 2(c)]. AA-1 and AA-3 stackings have the same energy, and both are the lowest energy of the AA stackings. However, AA-0 stacking exhibits the highest energy. The mutual transformation between AA-1 and AA-3 stackings only needs to overcome the 11.39 meV energy barrier. Among 2H phase AB stackings, the energy of AB-1 stacking is the lowest, and the AB-3 tacking is only 0.81 meV higher than that of AB-1. AB-1 stacking also only requires to overcoming a barrier of 12.24 meV to transform AB-3 stacking. We find an interesting phenomenon that the 2H phase overcomes the energy barrier less than half of the 1T phase from AA-1 (AB-1) stacking sliding to AA-3 (AB-3) stacking. Since the interlayer interaction of the 2H phase is weaker than that of the 1T phase, it can be verified by the ELF [see Figure S5(b, e)]. The above analysis indicates that AA-0, AB-1, AB-3 stackings, and AA-1, AA-3, AB-1 stackings are stable for 1T and 2H phase OsBr$_2$ bilayer, respectively. In these typical stackings, the AB-1, AB-3 stackings of 1T phase and AA-1, AA-3 stackings of 2H phase break the xy mirror symmetry. It is well known that the breaking of xy mirror symmetry could induce out-of-plane ferroelectric polarization ($\emph{P}$) \cite{58}. To confirm this, we calculated the plane averaged electrostatic potential, as shown in Figure S8 and Figure S9. The electrostatic potential difference are 63.85 (-63.85) meV and 16.26 (-16.26) meV for 1T phase AB-1 (AB-3) stacking, and 2H phase AA-1 (AA-3) stacking, respectively. Simultaneously, they have the same energy, which shows that 1T phase AB-1 (AB-3) stacking, and 2H phase AA-1 (AA-3) stacking have out-of-plane $\emph{P}$. The sliding displacement are shown in Figure S10, the $\emph{P}$ value of 1T phase AB-1 (AB-3) stacking, and 2H phase AA-1 (AA-3) stacking are up to 8.05$\times$10$^{-13}$ C/m and 2.29$\times$10$^{-13}$ C/m, respectively.

Whether there are novel physical properties in other atypical stacking orders? To reply to the question, we calculated the total energy of 1T phase and interlayer exchange energy of 2H phase stacking dependent for the entire 2D space of lateral shifts. As shown in Figure 2(b, f), AA-0, AB-1, and AB-3 stackings still have the global minimum for 1T phase AA and AB stackings. Moreover, the double barrier wells can be clearly observed in the AB stacking, which is a typical characteristic of ferroelectricity. Figure 2(d, h) shows 2H phase interlayer interaction replying on lateral shift. The interlayer exchange energy is defined as the difference in energy between AFM and FM interlayer spin configurations. We find that both AA and AB stackings have AFM coupling between interlayers. In addition, the interlayer AFM coupling of AA-1, AA-3, and AB-1 stackings is significantly enhanced, which is different from the GdI$_2$ interlayer FM coupling enhancement \cite{20}. It is origin from 2H phase OsBr$_2$ and GdI$_2$ occupy different orbits. In the GdI$_2$, FM interlayer coupling is beneficial to reduce total energy, while AFM interlayer coupling is advantageous to the stability of the OsBr$_2$. It also confirms from another angle that AA-1, AA-3, and AB-1 stackings are the lowest energy configurations of AA and AB stackings, respectively.

\subsection{Stacking-Dependent Ferrovalley Property of Bilayer OsBr$_2$}
%%%%%%%%%%%%%%%%%%%%%%%%%%%%%%%%%%%%%%%%%%%%%%%%%%%%%%%%%%%%%%%%%%
\begin{figure}[htb]
\begin{center}
\includegraphics[angle=0,width=1.0\linewidth]{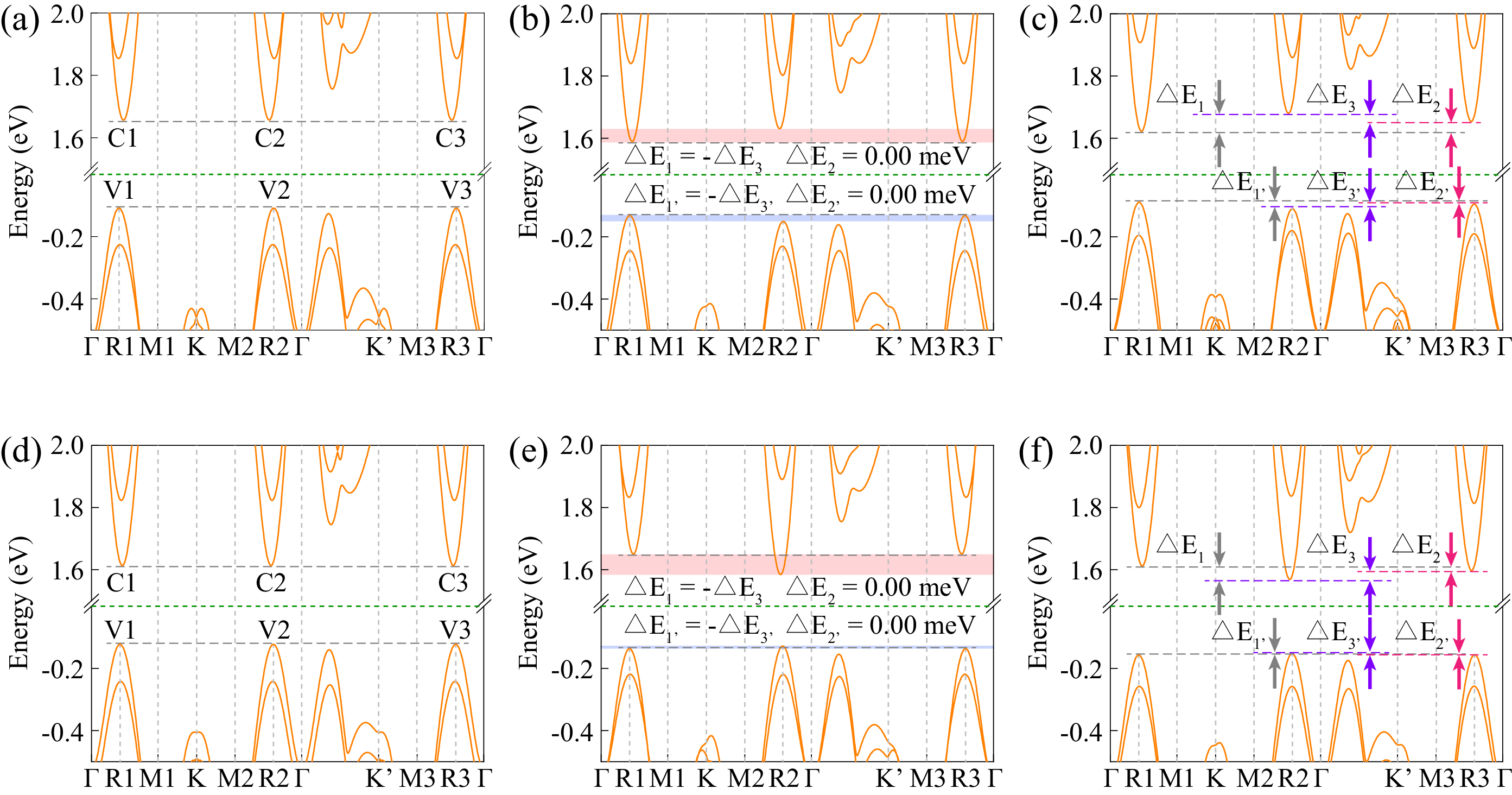}
\caption{
Band structures of 1T phase bilayer OsBr$_2$ without considering the SOC. (a-f) are AA-0, AA-4, AA-6, AB-0, AB-4, and AB-6 stacking, respectively. The valence band maximum and conduction band minimum of R1, R2, and R3 are named V1, V2, V3, C1, C2, and C3, respectively. The $\Delta E_1$, $\Delta E_{1'}$, $\Delta E_2$, $\Delta E_{2'}$, $\Delta E_3$, and $\Delta E_{3'}$ are defined as E$_{\rm C1}$-E$_{\rm C2}$, E$_{\rm C1}$-E$_{\rm C3}$, E$_{\rm C2}$-E$_{\rm C3}$, E$_{\rm V1}$-E$_{\rm V2}$, E$_{\rm V1}$-E$_{\rm V3}$, and E$_{\rm V2}$-E$_{\rm V3}$, respectively. The valence and conduction band valley splitting are indicated by the light blue and pink shading, respectively.
}
\end{center}
\end{figure}
%%%%%%%%%%%%%%%%%%%%%%%%%%%%%%%%%%%%%%%%%%%%%%%%%%%%%%%%%%%%%%%%%%

Stacking orders not only changes the interlayer coupling strength and induces out-of-plane ferroelectric polarization, but also contributes rich valley physics. As shown in Figure 3, the band structures of the 1T phase OsBr$_2$ bilayer without the SOC effect are investigated. It can be found that the valence and conduction bands of R1, R2, and R3 valleys are degenerate for the AA-0 and AB-0 stackings [see Figure 3(a, d)], since the coexistence of C$_{\rm 3z}$ rotational symmetry and xy plane mirror symmetry between the lower and upper layers. The AA-4 and AB-4 stackings are obtained by sliding t$_{//}$[$\frac{1}{2}$, 0, 0] of AA-0 and AB-0 stackings, respectively. The rotational symmetry reduces from C$_{\rm 3z}$ to C$_{\rm 2z}$ for AA-4 and AB-4 stackings, simultaneously, the xy plane mirror symmetry breaks. Intriguingly, the valley degeneracy of R1 (R3) and R2 points disappears, as shown in Figure 3(b, e) while the valley of R1 and R3 points are still degenerate. The valley polarization of AA-4 and AB-4 stackings are up to 65.70 meV and 74.28 meV (see Table SI), respectively. Different from the current mainstream VSe$_2$ \cite{13}, GdI$_2$ \cite{20}, and CrOBr \cite{22} originate from the combination of $\hat{T}$ symmetry breaking and SOC effect, which is completely caused by lattice symmetry breaking. Importantly, the lattice symmetry breaking induces valley polarization that is of the same order as theirs, and also has the conditions that can be observed at room temperature. Ulteriorly, the AA-4 and AB-4 stackings slide t$_{//}$[-$\frac{1}{10}$, $\frac{1}{10}$, 0] to acquire the AA-6 and AB-6 stackings, respectively. The rotation symmetry of C$_{\rm 2z}$ is further decreased to C$_{\rm 1}$ for the AA-6 and AB-6 stackings. The valley degeneracy of R1 and R3 points also vanishes, as shown in Figure 2(c, f), the tri-state valley polarization exhibits. It has never been reported in previous studies. We define the valley polarization of conduction and valence bands as $\Delta E_1$, $\Delta E_2$, $\Delta E_3$, and $\Delta E_{1'}$, $\Delta E_{2'}$, $\Delta E_{3'}$, respectively. When the SOC is switched on, as shown in Figure S11, the valley degeneracy of AB-0 stacking is still retained. For the AB-4 stacking, the valley polarizations of valence and conduction bands decreased from 66.08 meV and 8.20 meV to 61.71 meV and 6.00 meV, respectively. In a word, the effect of SOC on valley polarization is very small. Therefore, the SOC effect can be ignored for valley polarization of 1T phase OsBr$_2$ bilayer.

%%%%%%%%%%%%%%%%%%%%%%%%%%%%%%%%%%%%%%%%%%%%%%%%%%%%%%%%%%%%%%%%%%
\begin{figure}[htb]
\begin{center}
\includegraphics[angle=0,width=0.8\linewidth]{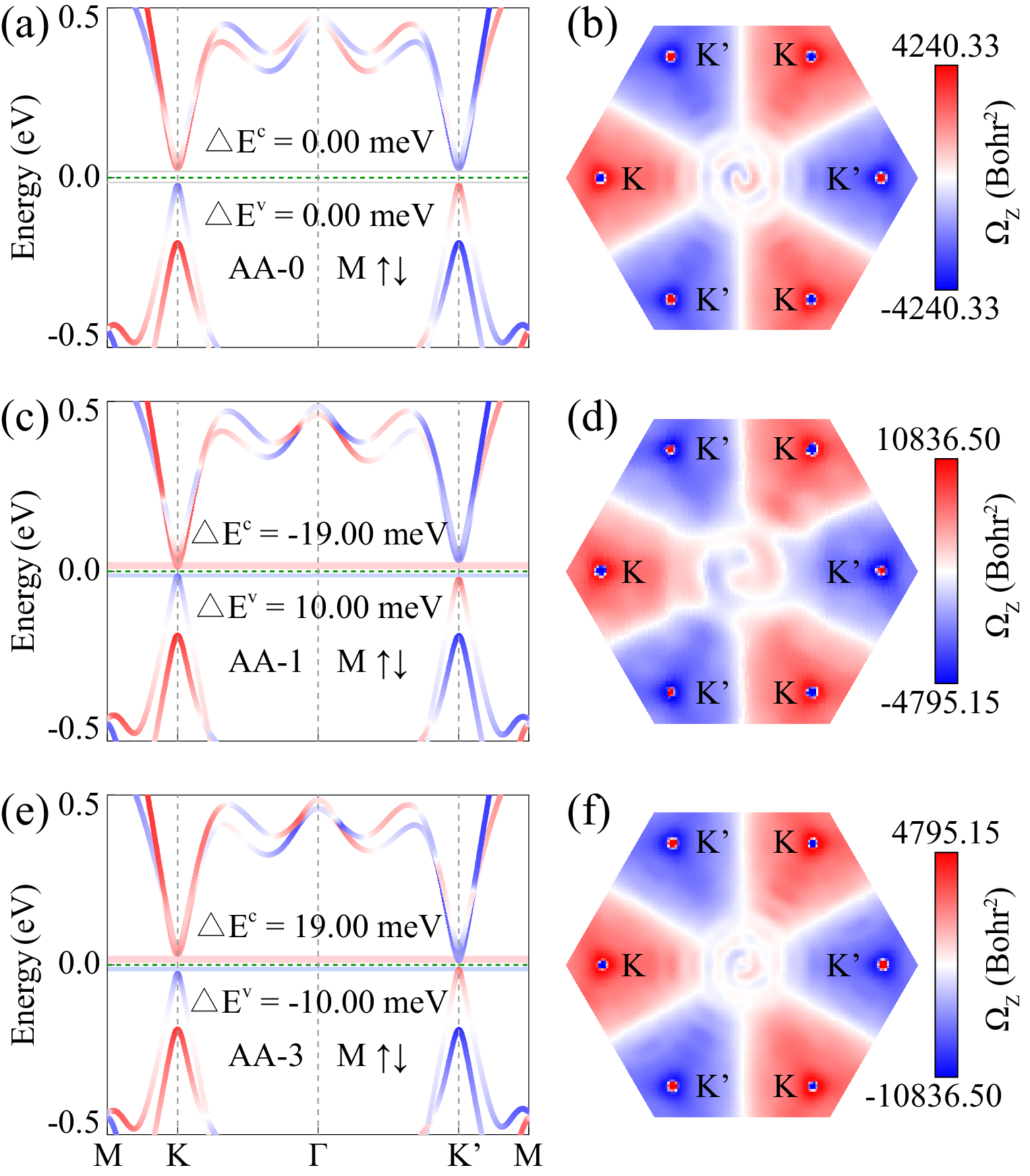}
\caption{
Spin-resolved band structures and Berry curvatures of 2H AA-0 (a, b), AA-1 (c, d), and AA-3 (e, f) stacking bilayer OsBr$_2$ under the SOC effect. The valence and conduction bands valley splitting are indicated by the light blue and pink shading, respectively.
}
\end{center}
\end{figure}
%%%%%%%%%%%%%%%%%%%%%%%%%%%%%%%%%%%%%%%%%%%%%%%%%%%%%%%%%%%%%%%%%%

For the 2H phase, OsBr$_2$ monolayer exhibits ferromagnetism due to orbital occupation, while OsBr$_2$ bilayer is more inclined to AFM coupling. Importantly, the MAE of 2H phase OsBr$_2$ bilayer is twice as enhanced as that of monolayer (see Table SII), which the easy magnetic direction along the z axis is the basis for investigating valley polarization. The valley polarization of the 2H phase derives from the combination of the $\hat{T}$ symmetry breaking and SOC effect. As shown in Figure S12(a), spin up and spin down bands are degenerate for AA stacking, meanwhile, the VBM and CBM at K and K' points are also degenerate in the absence of SOC. When the SOC is included, as shown in Figure 4(a), the degeneracy of spin up and spin down bands vanishes. We are mainly concerned the changes in the valence band. The spin up band moves below the spin down band in the K valley, while the spin up moves above the spin down band in the K' valley. More intriguingly, the spin up and spin down bands of K and K' valleys in energy are degenerate. In addition, as shown in Figure 4(b), the Berry curvatures of K and K' valleys have opposite signs and equal magnitudes (-4240.33 Bohr$^2$ for the K point and 4240.33 Bohr$^2$ for the K' point).

%%%%%%%%%%%%%%%%%%%%%%%%%%%%%%%%%%%%%%%%%%%%%%%%%%%%%%%%%%%%%%%%%%
\begin{figure}[htb]
\begin{center}
\includegraphics[angle=0,width=1.0\linewidth]{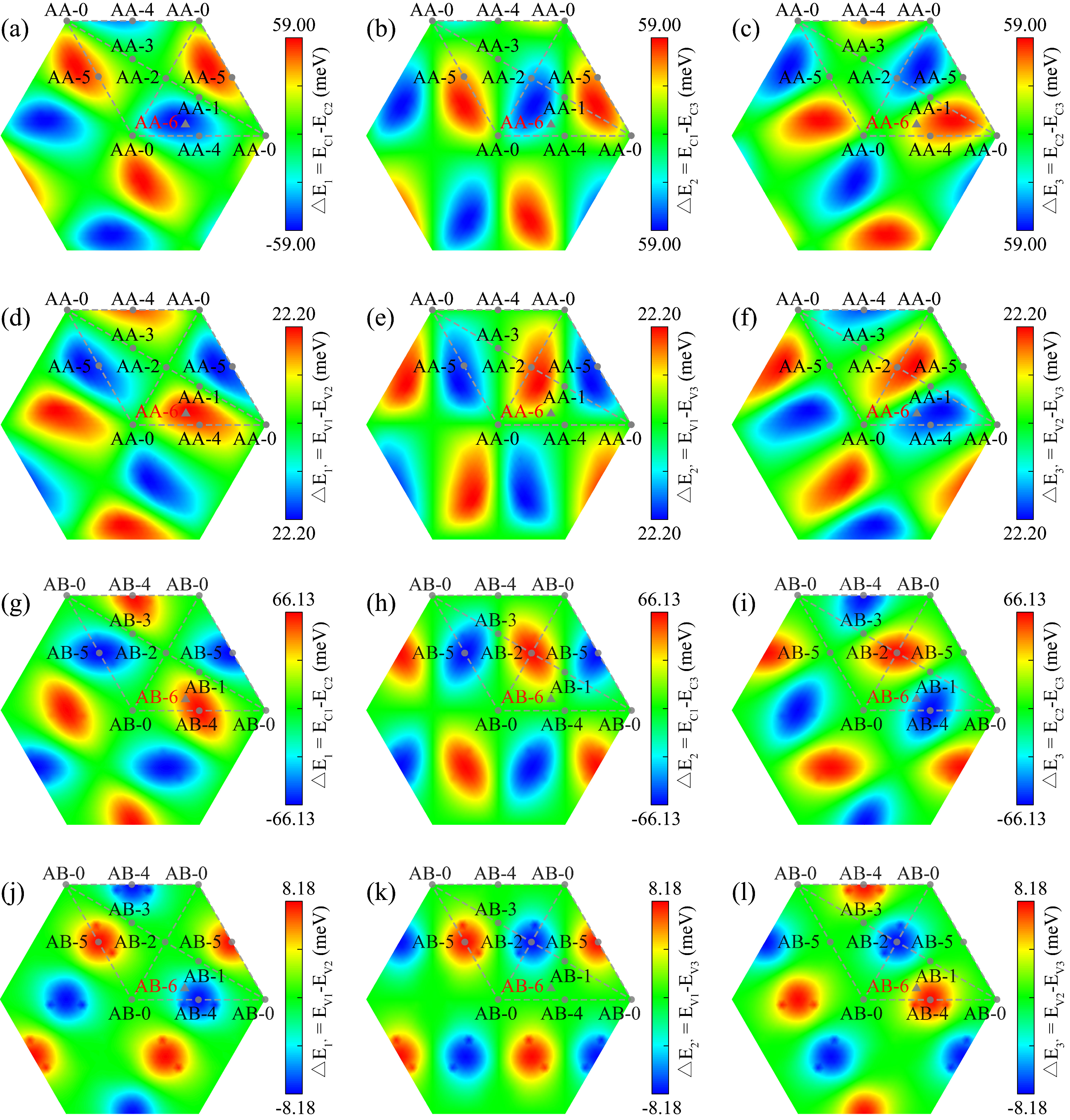}
\caption{
The valley polarization of 1T phase bilayer OsBr$_2$ in full space. (a-c) and (d-f) are valley polarization of AA stacking conduction band [(a) E$_{\rm C1}$-E$_{\rm C2}$, (b) E$_{\rm C1}$-E$_{\rm C3}$, and (c) E$_{\rm C2}$-E$_{\rm C3}$] and valence band [(d) E$_{\rm V1}$-E$_{\rm V2}$, (e) E$_{\rm V1}$-E$_{\rm V3}$, and (f) E$_{\rm V2}$-E$_{\rm V3}$], respectively. (g-i) and (j-l) are valley polarization of AB stacking conduction band [(a) E$_{\rm C1}$-E$_{\rm C2}$, (b) E$_{\rm C1}$-E$_{\rm C3}$, and (c) E$_{\rm C2}$-E$_{\rm C3}$] and valence band [(d) E$_{\rm V1}$-E$_{\rm V2}$, (e) E$_{\rm V1}$-E$_{\rm V3}$, and (f) E$_{\rm V2}$-E$_{\rm V3}$], respectively.
}
\end{center}
\end{figure}
%%%%%%%%%%%%%%%%%%%%%%%%%%%%%%%%%%%%%%%%%%%%%%%%%%%%%%%%%%%%%%%%%%

When the AA-0 stacking slips to AA-1 stacking, the degeneracy of spin up and spin down bands vanishes without the SOC effect due to reduce the lattice symmetry [see Figure S12(b)]. When the SOC is switched on, as shown in Figure 4(c), the valence and conduction bands produce 10.00 meV and -19.00 meV valley polarization between the K and K' points, respectively. It means that the interlayer slipping reduces the lattice symmetry and induces the out-of-plane ferroelectric polarization, resulting in the valley polarization. On account of the valley polarization production, as shown in Figure 4(d), although the Berry curvature of K and K' points are still opposite, they are no longer equivalent (-4795.15 Bohr$^2$ for the K point and 10836.50 Bohr$^2$ for the K' point). Further sliding to AA-3 stacking, as shown in Figure S12(c) and Figure 4(e), the valley polarization is reversed by manipulating the direction of ferroelectric polarization. Simultaneously, the Berry curvatures of K and K' points are also reversed to 4795.15 and 10836.50 Bohr$^2$, respectively. Hence, ferroelectric polarization and valley polarization can be effectively regulated by interlayer sliding. In addition, it should be noted that the spin up and spin down energy bands of the AB stackings are always degenerate, as shown in Figure S13, since all AB stackings have $\hat{P}$ symmetry. Importantly, the AB stackings have a greater valley polarization, which the total valley polarization is up to $\thicksim$ 177 meV.

%%%%%%%%%%%%%%%%%%%%%%%%%%%%%%%%%%%%%%%%%%%%%%%%%%%%%%%%%%%%%%%%%%
\begin{figure}[htb]
\begin{center}
\includegraphics[angle=0,width=1.0\linewidth]{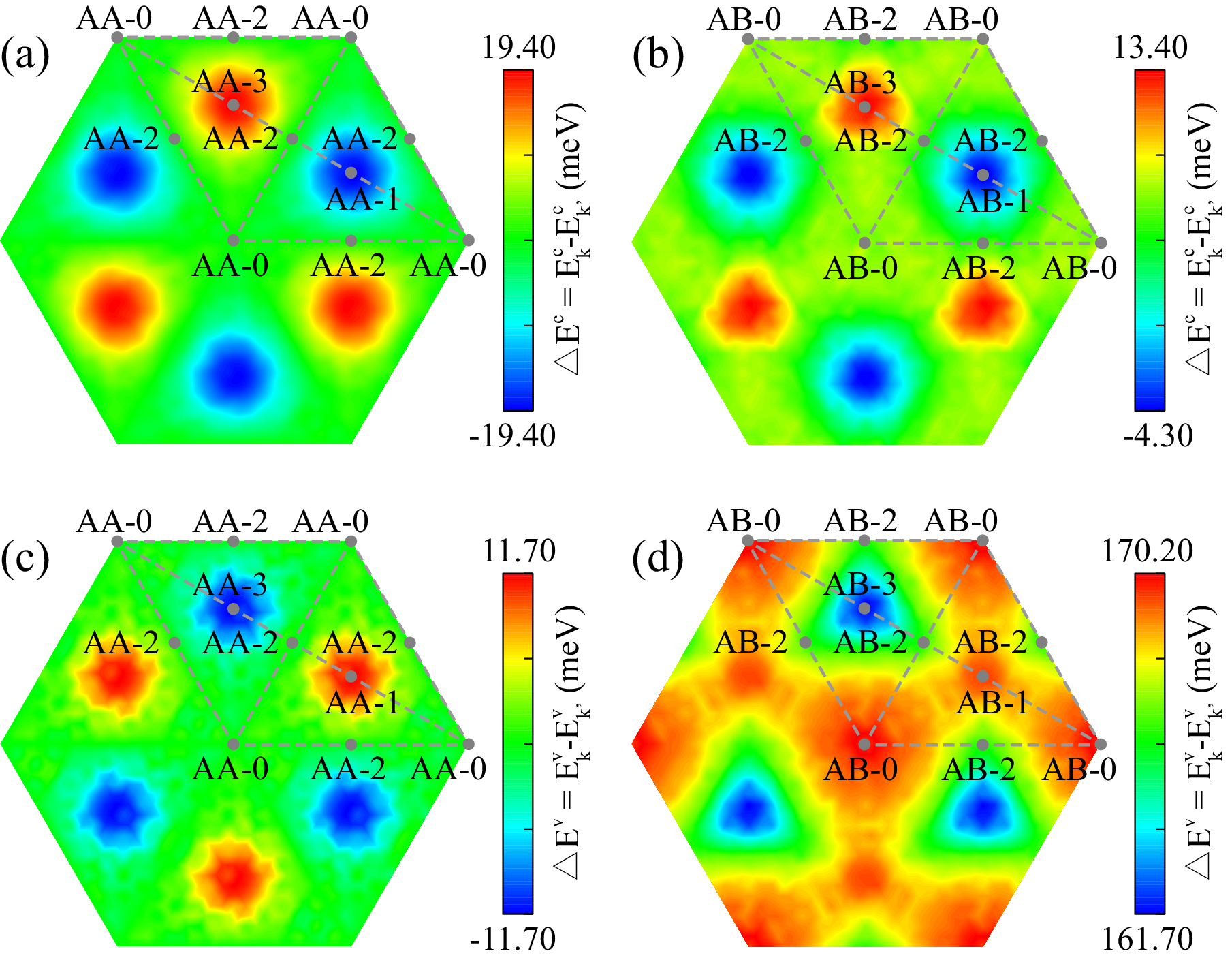}
\caption{
The valley polarization of 2H phase bilayer OsBr$_2$ in full space. (a) and (b) are valley polarization of AA and AB stacking conduction band, respectively. (c) and (d) are valley polarization of AA and AB stacking valence bands, respectively.
}
\end{center}
\end{figure}
%%%%%%%%%%%%%%%%%%%%%%%%%%%%%%%%%%%%%%%%%%%%%%%%%%%%%%%%%%%%%%%%%%

For a more comprehensive understanding of the details of valley polarization, we calculated the valley polarization of 1T and 2H phase for the full 2D space of lateral shifts, as shown in Figure 5 and Figure 6. Figure 5(a-f) and Figure 5(g-l) exhibit the tri-state valley polarization of 1T phase AA and AB stackings, respectively. The valley polarization of AA-0 and AB-0 stackings is 0 meV, since they have the C$_{\rm 3z}$ rotational symmetry, xy plane mirror symmetry and $\hat{P}$ symmetry, respectively. Interestingly, whether it's AA and AB stackings or valence and conduction bands, the valley polarization has C$_{\rm 2z}$ rotational symmetry. It is originated from the 1T phase with C$_{\rm 2z}$ rotational symmetry during the sliding process. For the AA stackings, the maximum and minimum values of valley polarization do not appear in highly symmetric structures. Besides, the valley polarization of valence and conduction bands show strong anisotropy. On the contrary, the maximum and minimum valley polarization values of AB stackings emerge in AB-2, AB-4, and AB-5 stackings with high symmetry structures. More intriguingly, the valley polarization of valence band exhibits isotropy, while the valley polarization of conduction band shows strong anisotropy.

In the following, we investigate the distribution of 2H phase valley polarization in the whole 2D space, as shown in Figure 6. The valley polarization of 2H phase exhibits C$_{\rm 3z}$ rotation symmetry due to the configuration with the C$_{\rm 3z}$ rotation symmetry. Furthermore, the valence and conduction bands of AA stackings and the conduction band of AB stackings appear isotropy, while the valence band of AB stackings emerges weak anisotropy. The maximum and minimum values of valley polarization of AA and AB stackings appear in AA-1 and AA-3 stackings, AB-1 and AB-3 stackings. Interestingly, both valence and conduction bands valley polarization of AA stackings are always equal magnitudes and opposite signs, while the valley polarization of conduction band is opposite signs and varies in value for the AB stackings, and the valley polarization of valence band is same opposite symbols and unequal magnitudes. It stems from the AA stackings have out-of-plane ferroelectric polarization, and the AA-1 and AA-3 stackings polarization values are equal in magnitude and opposite in sign. Therefore, the external electric field can effectively manipulate the ferroelectric polarization and induce the switching of valley polarization \cite{59}.

For the 1T phase OsBr$_2$, the valley polarization is entirely caused by symmetry breaking during the sliding process. As listed in Table SIII, when the system has a symmetry higher than C$_{3v}$, there will be no spontaneous valley polarization. However, when the C$_3$ symmetry is broken during the slip process, the valley polarization is induced. Interestingly, this rule does not apply to the 2H phase OsBr$_2$ bilayer, as listed in Table SIV. The valley polarization of 2H phase originates from the breaking $\hat{T}$ symmetry and SOC effect. The monolayer already exhibits spontaneous valley polarization. In order to make the bilayer valley polarization disappear, it is only possible to realize in the AA stackings due to the M$_{xy}$ mirror symmetry, as shown in Figure 6.

Based on this, it is expected to realize a large number of non-volatile multiferroic devices, such as, electrically controlled polarizer \cite{60}, and electronically controlled valley polarization device \cite{61}, and so on. Here, we mainly introduce the concept of the electronically controlled polarizer. The device can selectively generate x or y-polarized light, and through an additional quarter-wave plate obtain circularly polarized light with specific chirality. The operation is reversible by manipulating the electric field. This new electronically controlled polarizer can accurately and quickly control polarization by electrical means, and can realize the regulation of valley polarization.

\subsection{Phase Transition Between 1T and 2H Phases OsBr$_2$}
%%%%%%%%%%%%%%%%%%%%%%%%%%%%%%%%%%%%%%%%%%%%%%%%%%%%%%%%%%%%%%%%%%
\begin{figure}[htb]
\begin{center}
\includegraphics[angle=0,width=1.0\linewidth]{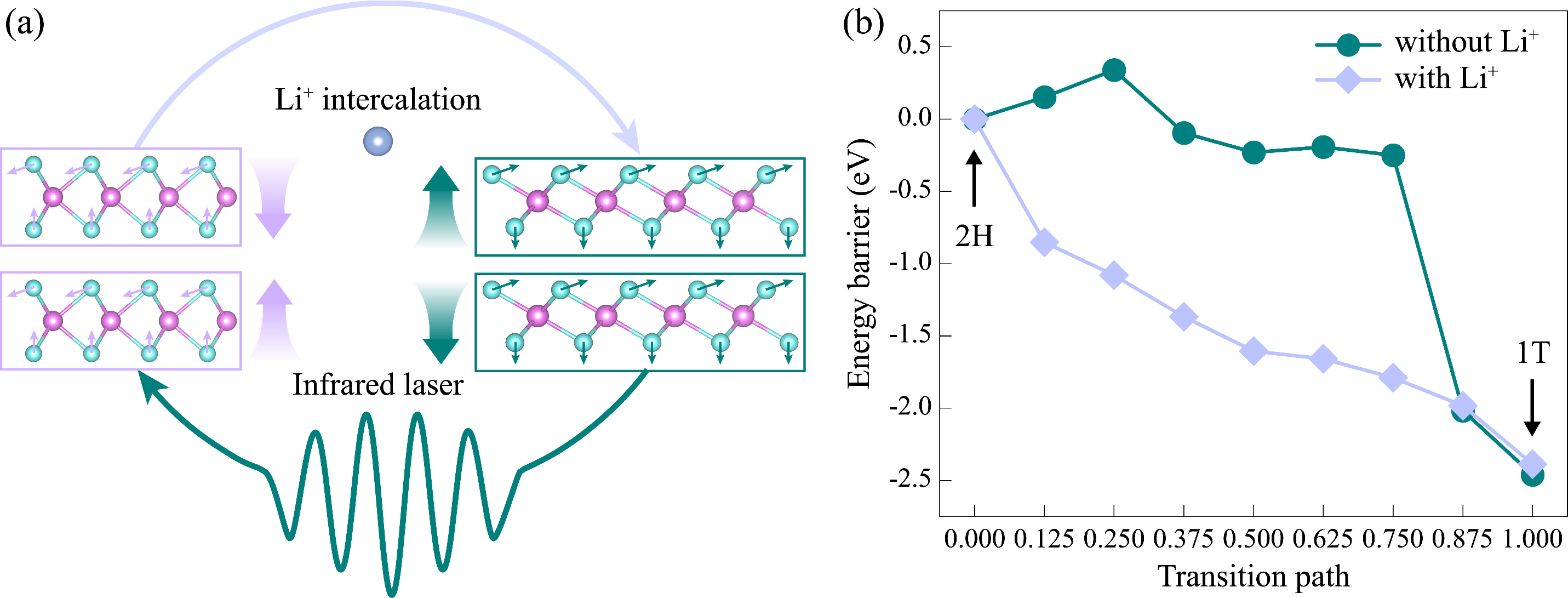}
\caption{
(a) Schematic diagram of phase transitions between 1T and 2H phases bilayer OsBr$_2$. (b) The energy barrier from 2H phase to 1T phase transition. The light blue and dark green lines indicate the presence or absence of Li$^+$ ion intercalation.
}
\end{center}
\end{figure}
%%%%%%%%%%%%%%%%%%%%%%%%%%%%%%%%%%%%%%%%%%%%%%%%%%%%%%%%%%%%%%%%%%

In the following, we focus on how to realize multiple types of valley polarization in OsBr$_2$ system. As shown in Figure 7(a), we propose to induce the transformation of 2H phase OsBr$_2$ into 1T phase by Li$^+$ ion intercalation. Then, the structure phase transition occurs from 1T phase into 2H phase by an infrared laser. To determine the feasibility of the scheme, we calculate the energy barrier from 2H phase to 1T phase by the nudged elastic band method. Firstly, in order to prove that the Li$^+$ intercalation structure is stable, we calculate the binding energy. It defined as E$_b$ = E$_{tot}$ - E$_{bilayer-OsBr2}$ - $\mu$$_{Li}$, where the E$_{tot}$, E$_{bilayer-OsBr2}$, and $\mu$$_{Li}$ are the total energies of Li$^+$ intercalated bilayer OsBr$_2$, the bilayer OsBr$_2$, and single isolated Li atom, respectively. The calculated E$_b$ is -4.967 eV, which indicates that the structure of Li$^+$ intercalated bilayer OsBr$_2$ is stable. Secondly, in the calculation, we intercalated a Li ion in the 1$\times$1 unit cell. Then, we consider three configurations of Li$^+$ intercalation, i.e., hollow, top-Os, and top-Br, as shown in Figure S14. As listed in Table SV, the top-Os configuration is the most stable. Therefore, we study the energy barrier of phase transitions based on top-Os configuration. The energy barrier is 339.96 meV for bilayer OsBr$_2$ from 2H phase to 1T phase. Interestingly, we find that the energy barrier will disappear from 2H phase to 1T phase when there is Li$^{+}$ ion intercalation. The reason is that Li$^{+}$ ion intercalation enhances the interlayer coupling and promotes the phase transition process, the resulting 2H to 1T phase transition has no energy barrier. In order to clearly understand the phase transition process, we marked the position change direction of Os and Br. Previous studies of transition metal dichalcogenides materials have also reached similar conclusions \cite{62,63,64}. In real experiments, the phase transition can be induced by introducing Li$^+$ intercalation in the following steps \cite{65}. Firstly, the Li$^+$ intercalation is realized by immersing of natural 2H phase crystals in butyllithium solution. Then, exfoliation is achieved immediately after this by ultrasonicating Li$_x$OsBr$_2$ in water. Finally, the samples are annealed at a desired temperature. Thus, the transformation from 2H phase to 1T phase is achieved. Moreover, the energy barrier for 1T phase transition into 2H phase can also be estimated by optical responses \cite{66}. The Gibbs free energy is
\begin{equation}
G_E(A, \omega) = F(\xi^0_A) - \frac{V \epsilon_0 \vec{E}\cdot\bar{\chi}(\xi^0_A, \omega)\cdot\vec{E}}{4}
\end{equation}
where the F($\xi^0_A$), V, $\epsilon_0$, $\vec{E}$, $\bar{\chi}(\xi^0_A, \omega)$ are the internal Helmohltz free energy, the volume of the formula unit, the vacuum permittivity, the optical alternating electric field, and the real part of susceptibility. This can be done experimentally with the following steps \cite{37}. The exfoliated OsBr$_2$ thin films on the DVD discs are prepared by drop-casting of the exfoliated OsBr$_2$ dispersion onto the top-side of DVD discs and drying at ambient temperature. The films are irradiated in air with the near-IR laser inside the commercial LightScribe DVD optical drive. For films with thicknesses of few microns, the 1T-2H phase reversion was complete after one write cycle.

\section{CONCLUSION}
In conclusion, we propose a mechanism to achieve the coexistence triferroic orders and multiple types of valley polarization in 2D materials. The mechanism is verified in OsBr$_2$ by structural phase transition. The 1T phase OsBr$_2$ bilayer exhibits the tri-state valley polarization at the non-high symmetric R point, while the 2H phase shows the valley polarization at the K point. We reveal the origin of valley polarization by Hamiltonian and symmetry. Moreover, the valley polarization and ferroelectric polarization of 1T phase AB stackings and 2H phase AA stackings can be regulated by interlayer sliding. Importantly, we proved that the 2H and 1T phases can transform each other through Li$^+$ ion intercalation and infrared laser. Our findings not only provide an idea for the coexistence triferroic order, but also offer a strategy for realizing multiple types of valley polarization.

\section*{Supporting Information}
Supporting Information is available from the Wiley Online Library or from the author.

\section*{ACKNOWLEDGEMENTS}
This work is supported by the National Natural Science Foundation of China (Grants No. 12474238, and No. 12004295). P. Li also acknowledge supports from the China's Postdoctoral Science Foundation funded project (Grant No. 2022M722547), the Fundamental Research Funds for the Central Universities (xxj03202205), the Open Project of State Key Laboratory of Surface Physics (No. KF2024$\_$02), and the Open Project of State Key Laboratory of Silicon and Advanced Semiconductor Materials (No. SKL2024-10).

\section*{Conflict of Interest}
The authors declare no conflict of interest.

\section*{Keywords}
valley splitting, multiple types of valley polarization, triferroic, structural phase transition, sliding ferroelectricity

%%%%%%%%%%%%%%%%%%%%%%%%%%%%%%%%%%%%%%%%%%%%%%%%%%%%%%%%%%%%%%%%%%%%%%%%%%%%%%%%%%%%%%%%%%%%%%%%%%%%%%%%%%%%%%%%%%%%%%%%%%%%%%%%%%%%

\end{document}